\documentclass[sigconf]{acmart}
\AtBeginDocument{%
  }

\copyrightyear{2025} \acmYear{2025} \setcopyright{acmlicensed}\acmConference[SC '25]{The International Conference for High Performance Computing, Networking, Storage and Analysis}{November 16--21, 2025}{St Louis, MO, USA} \acmBooktitle{The International Conference for High Performance Computing, Networking, Storage and Analysis (SC '25), November 16--21, 2025, St Louis, MO, USA} \acmDOI{10.1145/3712285.3759850} \acmISBN{979-8-4007-1466-5/2025/11}

\usepackage{algorithmic}
\usepackage{graphicx}
\usepackage{textcomp}
\usepackage{xcolor}
\usepackage{booktabs} 
\usepackage{fancyhdr}
\usepackage{balance}
\usepackage[T1]{fontenc}
\usepackage{aecompl}
\usepackage{makecell}
\usepackage[version=4]{mhchem}  
\usepackage{breakurl,bm}
\usepackage{multirow}
\usepackage{multicol}
\usepackage{booktabs}
\usepackage[switch]{lineno}

\newcommand{\KP}[1]{{#1}}

\newcommand{\EQ}{\begin{equation}}
\newcommand{\EN}{\end{equation}}

\begin{document}

\title{Deep Learning-Enabled Supercritical Flame Simulation at Detailed Chemistry and Real-Fluid Accuracy Towards Trillion-Cell Scale}

\thanks{$\dag$ Zhuoqiang Guo and Runze Mao contributed equally to this work.}
\thanks{$*$ Corresponding authors: Guangming Tan ( tgm@ict.ac.cn ), Weile Jia \\ ( jiaweile@ict.ac.cn ), Zhi X. Chen ( chenzhi@pku.edu.cn ).}

\author{Zhuoqiang Guo\textsuperscript{$\dag$}}
\affiliation{
  \institution{State Key Lab of Processors, Institute of Computing Technology, Chinese Academy of Sciences}
  \city{}
  \country{}
}
\affiliation{
  \institution{University of Chinese Academy of Sciences}
  \city{Beijing}
  \country{China}
}
\email{guozhuoqiang20z@ict.ac.cn}

\author{Runze Mao\textsuperscript{$\dag$}}
\affiliation{
  \institution{State Key Lab of Turbulence and Complex Systems, College of Engineering, Peking University}
  \city{Beijing}
  \country{China}
}
\email{maorz1998@stu.pku.edu.cn}

\author{Lijun Liu}
\affiliation{
  \institution{Department of Mechanical Engineering, Graduate School of Engineering, Osaka University}
  \city{Osaka}
  \country{Japan}
}
\email{liu@mech.eng.osaka-u.ac.jp}

\author{Guangming Tan\textsuperscript{$*$}}
\affiliation{
  \institution{State Key Lab of Processors, Institute of Computing Technology, Chinese Academy of Sciences}
  \city{Beijing}
  \country{China}
}
\email{tgm@ict.ac.cn}

\author{Weile Jia\textsuperscript{$*$}}
\affiliation{
  \institution{State Key Lab of Processors, Institute of Computing Technology, Chinese Academy of Sciences}
  \city{Beijing}
  \country{China}
}
\email{jiaweile@ict.ac.cn}

\author{Zhi X.Chen\textsuperscript{$*$}}
\affiliation{
  \institution{State Key Lab of Turbulence and Complex Systems, College of Engineering, Peking University}
  \city{}
  \country{}
}
\affiliation{
  \institution{AI for Science Institute}
  \city{Beijing}
  \country{China}
}
\email{chenzhi@pku.edu.cn}


\renewcommand{\shorttitle}{Deep Learning-Enabled Supercritical Flame Simulation Towards Trillion-Cell Scale}
\renewcommand{\shortauthors}{Zhuoqiang Guo, Runze Mao, et al.}

\begin{abstract}

For decades, supercritical flame simulations incorporating detailed chemistry and real-fluid transport have been limited to millions of cells, constraining the resolved spatial and temporal scales of the physical system.
We optimize the supercritical flame simulation software DeepFlame---which incorporates deep neural networks while retaining the real-fluid mechanical and chemical accuracy---from three perspectives: parallel computing, computational efficiency, and I/O performance.
Our highly optimized DeepFlame achieves supercritical liquid oxygen/methane (LOX/\ce{CH4}) turbulent combustion simulation of up to 618 and 154 billion cells with unprecedented time-to-solution, attaining 439/1186 and 187/316 PFlop/s (32.3\%/21.8\% and 37.4\%/31.8\% of the peak) in FP32/mixed-FP16 precision on Sunway (98,304 nodes) and Fugaku (73,728 nodes) supercomputers, respectively.
This computational capability surpasses existing capacities by three orders of magnitude, enabling the first practical simulation of rocket engine combustion with >100 LOX/\ce{CH4} injectors. This breakthrough establishes high-fidelity supercritical flame modeling as a critical design tool for next-generation rocket propulsion and ultra-high energy density systems.

\end{abstract}




\maketitle




\section{Introduction}\label{sec:introduction}

\begin{figure}[htbp]
\centerline{\includegraphics[width=\columnwidth]{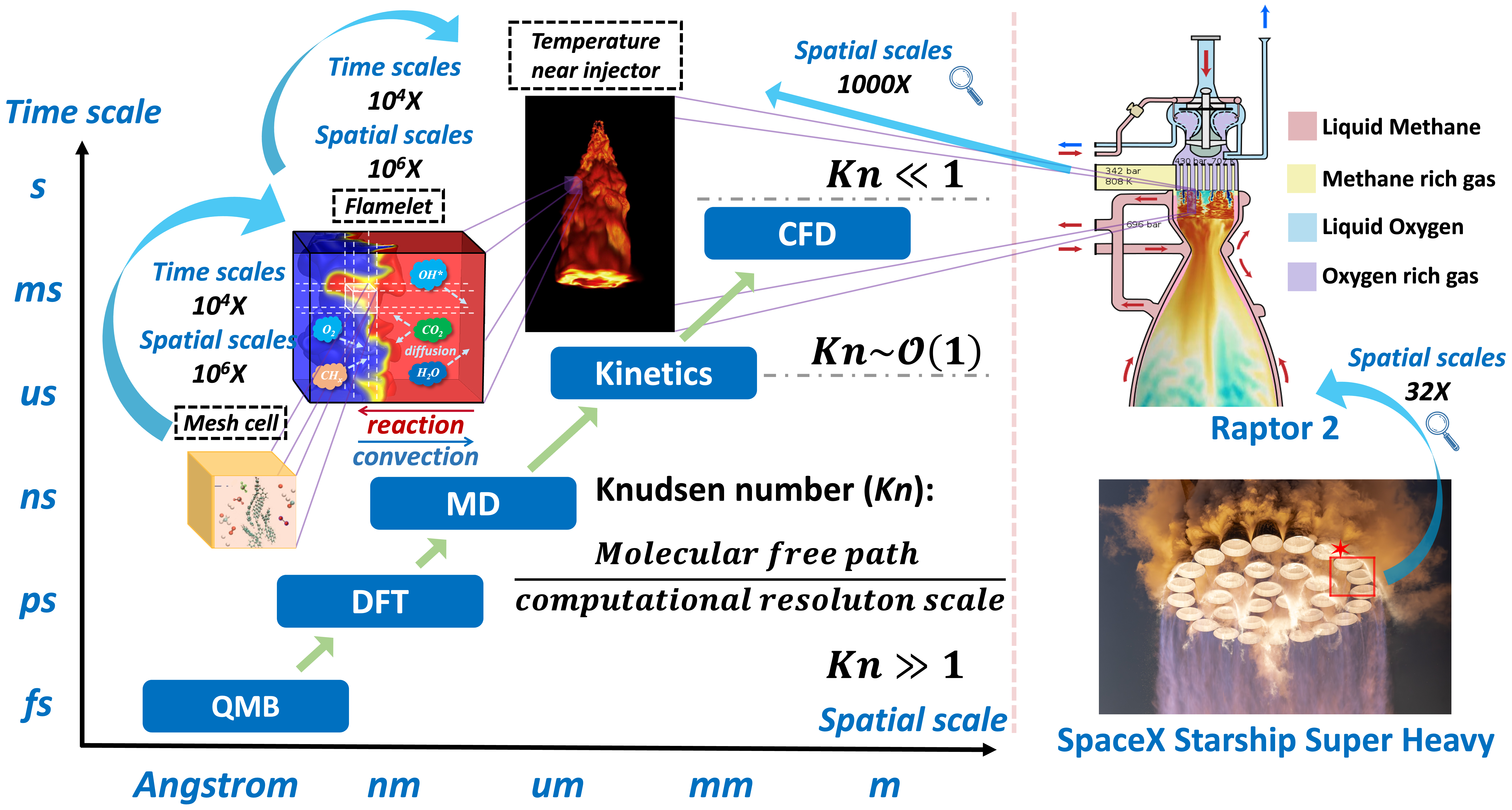}}
\caption{Overview of multi-scale modeling for rocket engine combustion.}
\label{fig:raptor2}
\end{figure}

Supercritical combustion, the process of burning fuel and oxidizer at pressures and temperatures above their critical points, is fundamental to a range of next-generation engineering systems. Its applications span from recyclable rocket engines, where fuels like methane are increasingly favored for their performance and reusability, to the development of high-efficiency power generation technologies utilizing supercritical carbon dioxide cycles.
In order to achieve ultimate engine performance, combustion enters extreme regimes of ultra-high pressure and temperature, resulting in unprecedented scientific and engineering challenges that hinder further development and maturing of the emerging technologies~\cite{jofre2021transcritical}. For example, as shown in Fig.~\ref{fig:raptor2}, SpaceX's \textit{Raptor} liquid oxygen/methane (LOX/CH$_4$) rocket engines powering the Starship operate at chamber conditions of 300 times the atmospheric pressure and over 3500~\textdegree C gas temperature \cite{wiki:raptor2}. Under such extreme conditions, thrust-generating fluids undergo a sequence of complex fluid mechanical and chemically reactive processes at pressure levels well above the thermodynamic critical points~\cite{yang2000modeling}. The dynamics of such so-called \textit{supercritical flames} are not only governed by the continuum-level Navier-Stokes (N-S) equations but also strongly influenced by the non-ideal (so-called \textit{real-fluid}) effects driven by molecular-level interactions~\cite{bellan2000supercritical}. 

High-fidelity simulation tools preserving sufficient physical and chemical details offer unique insights and predictive capabilities for device design and optimization at much lower economic and time costs~\cite{POINSOT20171}.
The major computational challenge for simulating supercritical flames is attributed to the complex chemistry and multi-physical processes taking place in turbulent reacting flows, which require highly resolved computational grids~\cite{domingo2023recent, pirozzoli2019high}. As shown in Fig.~\ref{fig:raptor2}, to simulate the flame behind a single fuel injector in a rocket engine, the spatial and temporal scales span over $\mathcal{O}(10^{12})$ (summed in three dimensions) and $\mathcal{O}(10^8)$, respectively. This results in a typical simulation of trillion-cell scale running over millions of time steps. 
In addition to the large grid size, another major difficulty stems from the so-called \textit{chemical reaction mechanism}, a meso/macroscopic network model (also see Fig.~\ref{fig:raptor2}) that describes the step-by-step molecular-level elementary reactions using chemical kinetics theory~\cite{glassman2014combustion}. To accurately capture the essential chemical reaction and transport processes in combustion engines, a \textit{detailed} mechanism including $\mathcal{O}(10^1)$ to $\mathcal{O}(10^2)$ chemical species (with $\sim5\times$ the number of reactions) is required \cite{lu2009toward}. Each species and related reaction steps add more degrees of freedom (DoF) to every grid cell, further increasing the computational complexity of the problem. 

Traditional algorithms, especially in large-scale computations, face issues such as high computational complexity and load imbalance. The development of artificial intelligence has brought new opportunities for accelerating reactive flow simulations. One such implementation is DeepFlame\cite{MAO2023108842}, an open-source framework implemented based on OpenFOAM, which uses AI algorithms to significantly speed up the reactive flow simulation process while maintaining almost the same level of accuracy. Moreover, the application of deep neural networks naturally addresses the load imbalance problem inherent in traditional algorithms, making it more suitable for large-scale simulations on modern exascale supercomputers.

However, DeepFlame still faces following three major challenges on Exascale supercomputers.

\textbf{The inability to utilize many-core supercomputers (millions of cores).} While leading systems like Sunway (39.9M cores) and Fugaku (7.6M cores) employ many-core architectures for peak performance, OpenFOAM's inadequate multi-threading support necessitates single-process-per-core execution. This disparity leads to prohibitive memory requirements and soaring communication overheads, ultimately undermining the performance benefits of modern supercomputing architectures.

\textbf{Low computational efficiency.} From an algorithmic perspective, the primary computations in DeepFlame consist of the DNN inference module for calculating chemical reactions and the PDE solving module for computational fluid dynamics. The DNN module primarily consists of dense computations, but faces performance challenges on modern Sunway and Fugaku architectures due to the absence of specialized accelerators for transcendental functions. This architectural limitation leads to significant inefficiencies in executing activation functions, hindering full utilization of their peak performance capabilities.
The PDE solving module involves classic sparse computations, suffering from poor locality, low computational density, computational dependencies, and write conflicts during parallelization. These issues result in extremely low computational efficiency for the PDE solving module.

\textbf{I/O bottlenecks during initialization phase.} Like most un-structured-mesh CFD codes, DeepFlame encounters critical I/O limitations when approaching trillion-cell simulations, with initialization data requirements exceeding 100 TB scale.

To address these challenges, we present optimizations for the DeepFlame software that enable efficient computing and scaling to exascale many-core architectures.
Our optimization involves four aspects: parallel strategy, PDE solving, DNN inference, and setup I/O. Our optimized code achieves 316.5PFlop/s (31.8\%) and 1.18EFlop/s (21.8 \%) mix-FP16 precision on Sunway and Fugaku, respectively. In addition, Our work makes it possible to perform 618 billion cells combustion simulations. 
Our key innovations are:
\begin{itemize}
\item A two-level parallelization scheme enable exploiting million-level cores of exascale supercomputers via process-distributed, thread-shared decomposition of unstructured cells. 
\item An effective many-core PDE solver based on mesh decomposition to fully harness the computational power of many-core architectures.
\item An efficient DNN inference implementation that maximizes the floating-point computational performance.
\item Three I/O optimization methods to solve the longstanding I/O bottleneck that has hindered large-scale simulations within the framework of OpenFOAM.
\end{itemize}

\section{Current State Of The Art}

As a unique branch of computational fluid dynamics (CFD), chemically reactive flow simulation with practical interests ranging from land/sea/air propulsion~\cite{POINSOT20171}, carbon-neutral power generation~\cite{kohse2021combustion} to astrophysical phenomena such as supernova explosion~\cite{poludnenko2019unified}, has been the focal research topic for a rapidly increasing number of investigations~\cite{domingo2023recent,mira2023hpc}. However, owing to the physical complexities and computational difficulties highlighted earlier, only in recent years few community efforts started to systematically assess the high-performance computing aspects of the existing codes for simulating turbulent reactive~\cite{ABDELSAMIE2021104935} and supercritical flows~\cite{ruiz2016numerical}.  

\begin{table*}[htb]
\centering
\caption{State-of-the-art performance of combustion simulations with detailed transport and chemistry. $^\dag$Cycle: characteristic flow time cycle, TGV and HIT follow their standard reference flow time definition. $^{**}$JET: slot jet cycle = slot width/jet velocity. \\ 
$^*$PF: planar premixed flame cycle = laminar flame thickness/speed. $^\S$ROK4E: a semi-implicit Runge Kutta method. 
$^\P$The only GPU counts in the table. $\ddag$Unreported system with Intel Xeon (E5-2698 v3) CPUs. 
$^{\dag\dag}$The baseline DeepFlame supercritical TGV implementation for the present work.
\label{table:sota}}
\resizebox{\textwidth}{!}{%
\begin{tabular}{c | c  c  c  c  c  c  c c c c} 
\hline
\hline
&Work & Year & EoS & \makecell[c]{Method \\ (fluid/chemistry)} & Benchmark & \makecell[c]{DoF} &  \#CPU/GPU & Machine & \makecell[c]{Time-to-Solution \\ (s/DoF/cycle$^\dag$)} & \makecell[c]{Flop/s \\ \% of peak} \\
\hline

\multirow{7}{*}{\rotatebox[origin=c]{90}{Ideal Gas}}& \makecell[l]{DINO \cite{abdelsamie2021taylor} }& 2021 & IG & \makecell[c]{E-FD / RK4 } & \makecell[c]{TGV/9-\ce{H2}} & 235M & 1K & SuperMUC-NG & 
$8.4\times10^{-6}$& 3.4T (3.36\%) \\

&\makecell[l]{YALES2 \cite{abdelsamie2021taylor} }& 2021 & IG & \makecell[c]{E-FV / CVODE } & \makecell[c]{TGV/9-\ce{H2}} & 235M & 0.8K & Irene J–Curie & 
$8.2\times10^{-6}$ & 2.3T (3.43\%) \\

&\makecell[l]{NEK5000 \cite{abdelsamie2021taylor} }& 2021 & IG & \makecell[c]{I-SE / CVODE } & \makecell[c]{TGV/9-\ce{H2}} & 235M & 0.6K & Piz Daint & 
$1.1\times10^{-5}$ & 2.4T (12.3\%) \\

&\makecell[l]{EBIFoam \cite{zirwes2023assessment}}& 2023 & IG & \makecell[c]{I-FV / CVODE } & \makecell[c]{TGV/9-\ce{H2}} & 235M & 0.5K & HoreKa & 
$4.7\times10^{-6}$ & --- \\

&\makecell[l]{S3D \cite{hawkes2012S3D}}& 2012 & IG & \makecell[c]{E-FD / RK4 } & \makecell[c]{JET$^{**}$/9-\ce{H2}} & 179B & 120K & Jaguar & ---  & --- \\

&\makecell[l]{PeleC \cite{henry2023pelec}}& 2023 & IG & \makecell[c]{E-FV / RK4 } & \makecell[c]{PF$^{*}$/21-\ce{CH$_4$}} & 4.16T & 27.6K$^\P$ & Summit & $1 \times 10^{-5}$  & --- \\

&\makecell[l]{DeepFlame \cite{MAO2023108842}}& 2023 & IG & \makecell[c]{I-FV / ODENet } & \makecell[c]{TGV/9-\ce{H2}} & 235M & 0.5K & Archer2 & $8.5\times10^{-7}$  & --- \\

\hline
\multirow{7}{*}{\rotatebox{90}{Supercritical}}&\makecell[l]{SiTCom-B~\cite{monnier2023numerical}}& 2023 & SRK & \makecell[c]{E-FV / RK4} & \makecell[c]{HIT/17-\ce{CH$_4$}} & 8M & --- & --- & ---  & --- \\

&\makecell[l]{CharlesX~\cite{chung2022interpretable}}& 2022 & PR & \makecell[c]{E-FV / ROK4E$^\S$ } & \makecell[c]{HIT/5-\ce{CH$_4$}} & 21M & 1K & Unknown$^\ddag$ & $1.2\times10^{-3}$  & --- \\


&\makecell[l]{Baseline \cite{xu2023detailed}$^{\dag\dag}$} & 2023 & PRNet & \makecell[c]{I-FV / ODENet } & \makecell[c]{TGV/17-\ce{CH$_4$}} & 46M & 1K & Fugaku &  $1.3\times10^{-4}$ & 13T (17.5\%) \\

\cline{2-11}

&\makecell[l]{our work (fp32) } & 2025 & PRNet & \makecell[c]{I-FV / ODENet } & \makecell[c]{TGV/17-\ce{CH$_4$}} & 3.4T & 3.5M & Fugaku & 
$8.5\times10^{-9}$
& \makecell{186.5P (37.4\%)} \\

&\makecell[l]{our work (fp32) } & 2025 & PRNet & \makecell[c]{I-FV / ODENet } & \makecell[c]{TGV/17-\ce{CH$_4$}} & 13.6T & 38.3M & Sunway & 
$3.2\times10^{-9}$& \makecell{438.9P (32.3\%)} \\ 

&\makecell[l]{our work (mix-fp16) } & 2025 & PRNet & \makecell[c]{I-FV / ODENet } & \makecell[c]{TGV/17-\ce{CH$_4$}} & 3.4T & 3.5M & Fugaku & 
$5.0\times10^{-9}$& \makecell{316.5P (31.8\%)} \\

&\makecell[l]{our work (mix-fp16) } & 2025 & PRNet & \makecell[c]{I-FV / ODENet } & \makecell[c]{TGV/17-\ce{CH$_4$}} & 13.6T & 38.3M & Sunway & 
$1.2\times10^{-9}$& \makecell{1.18E (21.8\%)} \\


\hline
\hline

\end{tabular}
}

\end{table*}

\textbf{Ideal-gas flame simulation}: On the basis of the well-established TGV benchmark for non-reactive CFD codes~\cite{wang2013high}, pioneering efforts were collected at the ``17th International Conference on Numerical Combustion" to build a standardized reactive TGV benchmark. Most of the state-of-the-art codes joined this campaign and those with published performance results~\cite{abdelsamie2021taylor} are listed in Table~\ref{table:sota}. 
Without the complication of real-fluid transport, the performance is mainly constrained by the fluid PDE spatiotemporal discretization and chemistry ODE integration methods. While different time marching (explicit/implicit) combined with finite difference (FD), finite volume (FV), spectral element (SE) schemes are implemented for DINO~\cite{abdelsamie2016towards}, YALES2~\cite{moureau2011large}, NEK5000~\cite{tomboulides1997numerical} codes, the Time-to-Solutions (ToSs) are similar, given 0.5$\sim$1K CPU cores used. This is because the ToS is primarily limited by the CVODE used for chemistry integration (except for DINO with explicit RK4). 
The CVODE method inherently exhibits load imbalance due to the spatial variability in chemical reaction rates across the computational domain. Regions with active reactions demand significantly smaller interval time steps due to high ODE stiffness, compared to areas devoid of reactions. This discrepancy, along with sparsity and MPI communication overhead for implicit CFD solvers, has limited the code scalability on large-scale platforms.

For ideal-gas combustion, as also listed in Table~\ref{table:sota}, there have been several large-scale simulations conducted using the S3D~\cite{chen2009terascale} and PeleC~\cite{henry2023pelec} codes. Both S3D and PeleC adopt explicit schemes for fluid and chemistry allowing good scalability attributes up to DoF of 179B~\cite{hawkes2012S3D} and 4.16T~\cite{henry2023pelec}, respectively, with the latter being a full-node run on Summit. Recently, PeleC is equipped with the implicit CVODE solver for more robust chemistry integration, at the cost of compensating scalability with the largest DoF so far up to 1.56T on Frontier~\cite{de2024pele}. Despite these advances, efficiently coupling fully implicit FV and CVODE at full-machine scale is still an unfulfilled task, even just for simple idea-gas situations.  

\textbf{Supercritical flame simulation}: 
The extreme complexity of supercritical flow physics has limited most investigations so far to two-dimensional (2D) simulations, seldom carrying detailed transport and chemistry accuracy. There have been few sophisticated codes available, such as RAPTOR~\cite{oefelein2005thermophysical}, AVBP~\cite{schmitt2011large}, CharlesX~\cite{ma2017entropy} and SiTCom-B~\cite{monnier2023numerical} as compared in~\cite{ma2017entropy,guven2019impact} showing similar predictive accuracy in 2D non-reactive benchmarks. 
Unfortunately, an established supercritical reactive benchmark is not yet available within the community. Two recent works, as listed in Table~\ref{table:sota}, adopted a reactive Homogeneous Isotropic Turbulence (HIT) configuration using the SiTComB~\cite{monnier2023numerical} and CharlesX~\cite{chung2022interpretable}, respectively. This supercritical HIT setup, however, involves random velocity field generation, which is not ideal for quantitative comparison of code accuracy and performance. To close this gap, in our earlier baseline work~\cite{xu2023detailed}, a supercritical reactive TGV benchmark was proposed, combining the ideal-gas TGV~\cite{abdelsamie2021taylor} with the supercritical HIT~\cite{chung2022interpretable} thermodynamic conditions. All of the above works were limited to small-scale DoF and MPI ranks. Considering the slow pace at which this particular research area is advancing, it would take decades to see a three-dimensional supercritical flame simulation of the practically relevant size of DoF with the conventional methods for detailed transport and chemistry.

\textbf{Deep learning approaches}: 
Recently, machine learning (ML) methodologies have emerged as a new paradigm for enhancing scientific computing~\cite{9355242,guo2022extending}, and the combustion field has also started to utilize machine learning approaches to improve simulation performance ( PINN~\cite{raissi2019physics}, NODE~\cite{saito2023data}, DFNN~\cite{milan2021deep}, etc.).
These advances strongly demonstrate that machine learning, especially deep learning methods, offer an accurate and efficient tool to approximate the complex phenomena in turbulent reactive flows~\cite{ihme2022combustion}. 
However, the existing studies are mostly at the proof-of-concept stage focusing on simplified 1D or 2D ``toy problems". Notably, the MMP~\cite{ding2021machine} and DeepFlame~\cite{zhang2024graphics} works are among the few attempts to deploy neural networks to accelerate simulation of lab-scale flames, but limited to workstation or local cluster scales. 

\begin{figure}[ht]
\centering
\includegraphics[width=\columnwidth]{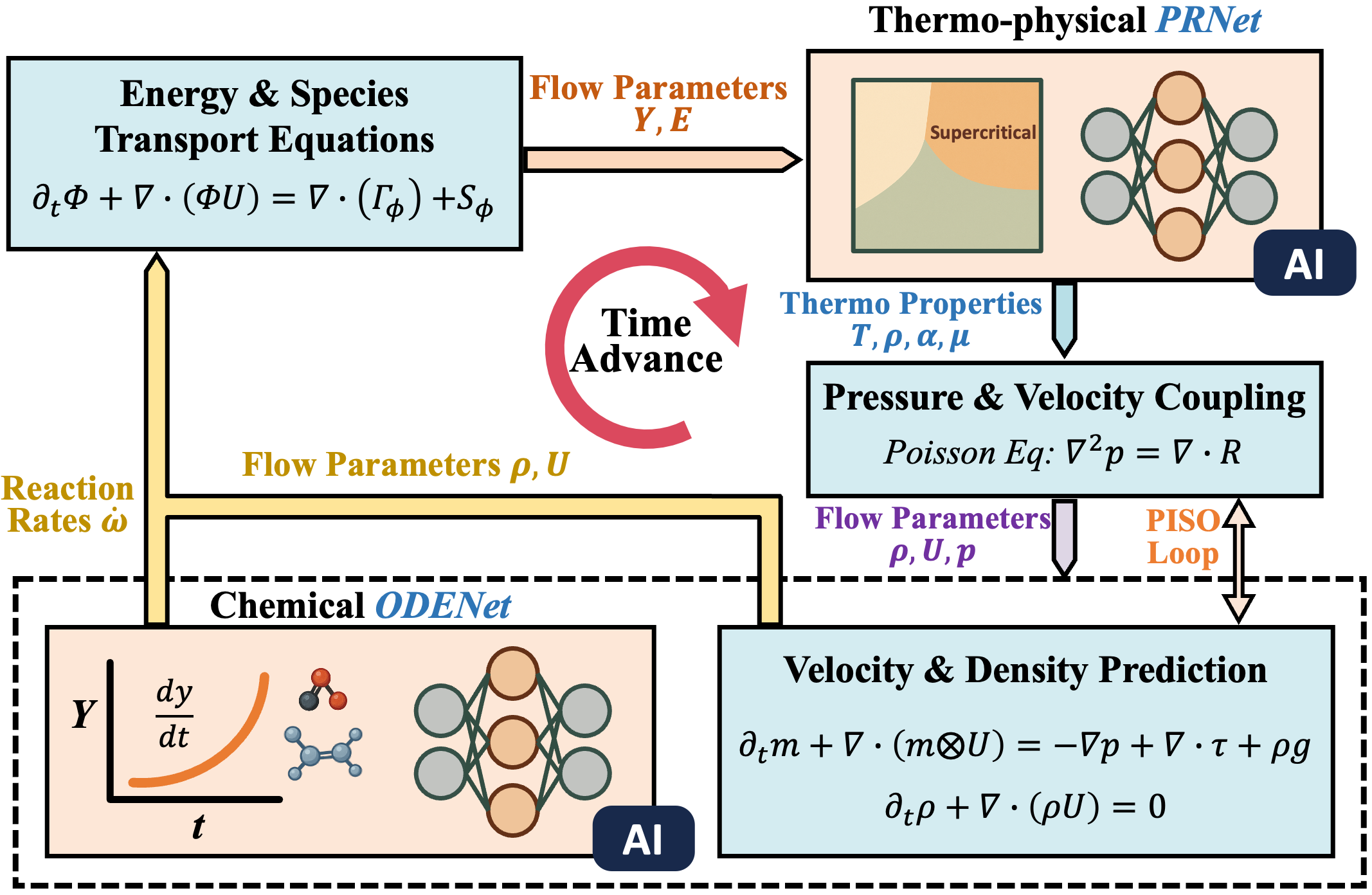}
\caption{the workflow of DeepFlame}
\label{fig:distribution_workflow}
\end{figure}

Of particular relevance to our work is the DeepFlame package~\cite{MAO2023108842}, which has been developed to facilitate robust integration of ML models for combustion simulations. DeepFlame is built on the open-source CFD platform OpenFOAM~\cite{jasak2007openfoam}. As shown in Fig.~\ref{fig:distribution_workflow}, DeepFlame follows the MPI-based parallel strategy of OpenFOAM, where the computational domain is discretized with an unstructured mesh and decomposed prior to runtime simulation. During time-marching (see Fig.~\ref{fig:distribution_workflow}), the transport PDEs (highlighted by blue blocks) are solved using a fully implicit FV method, while the chemical source terms are computed by inferencing a pre-trained neural network (NN) called ODENet. 
The ODENet takes the thermochemical state variables ($T$, $p$, $Y_i$) as the input and obtain the chemical source terms ($\dot\omega$). 

For simulations under supercritical conditions, an additional neural network, termed the PRNet, is employed to predict the real-fluid thermodynamic and transport properties based on Peng-Robinson EoS. The PRNet takes as input the flow field and thermodynamic state variables, including energy ($E$), pressure ($p$), and species mass fraction ($Y$), and outputs key supercritical mixture properties: density ($\rho$), temperature ($T$) and thermophysical properties such as viscosity ($\mu$) and thermal diffusivity ($\alpha$), etc.
The use of the above two NN models greatly reduces the computational complexity of the chemistry and transport calculations, which makes the time-to-solution of DeepFlame orders of magnitude faster than the other works with conventional methods in Table~\ref{table:sota}, for both ideal-gas~\cite{MAO2023108842} and supercritical~\cite{xu2023detailed} configurations.



\section{Innovations Realized}\label{sec:innovations}

\begin{figure}[ht]
    \centering
    \includegraphics[width=1\columnwidth]{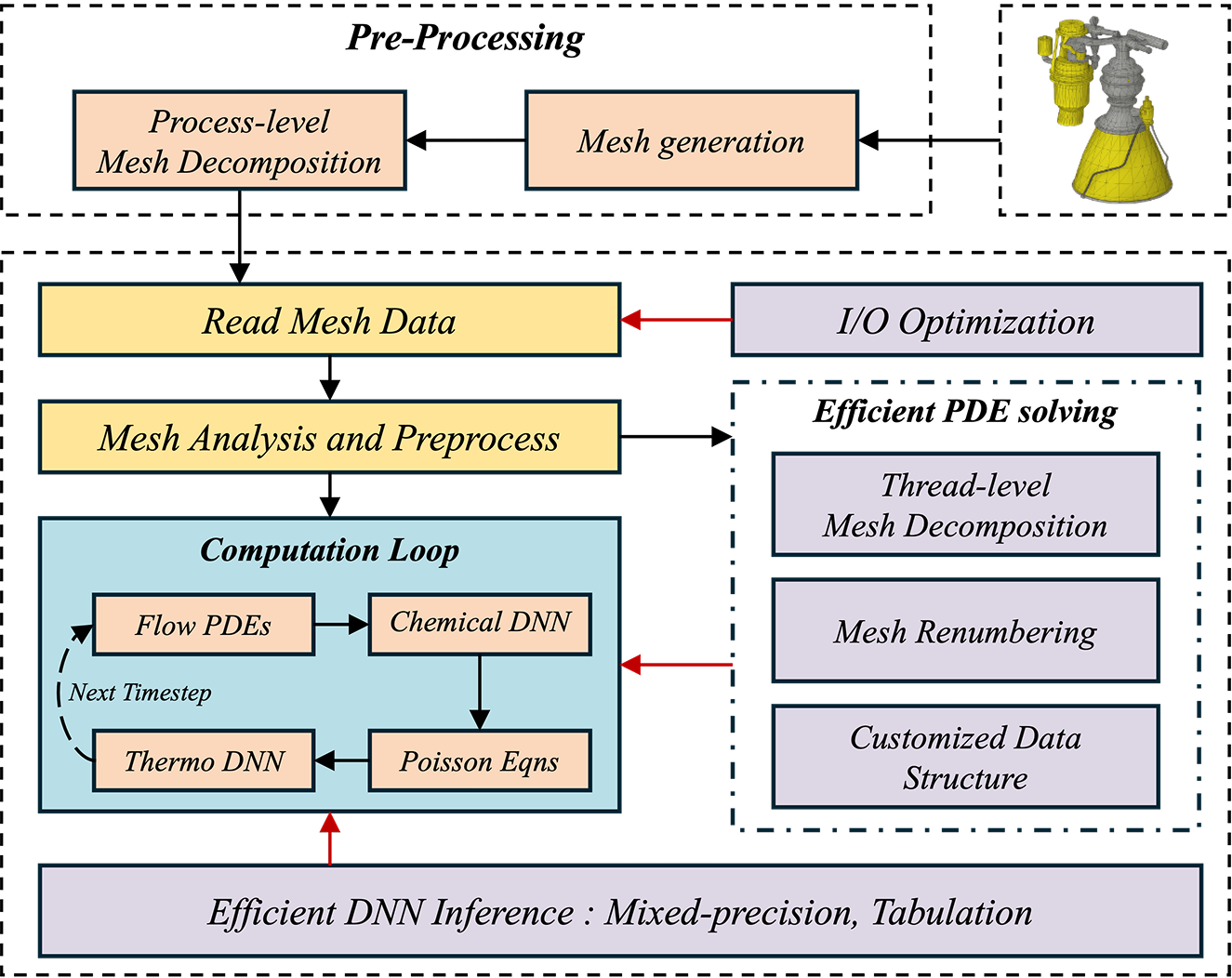}
    \caption{the workflow of DeepFlame and our optimization modules.}
    \label{fig:optimizations}
\end{figure}

This section will introduces our four contributions, including the two-level parallelization scheme, a many-core PDE solver, DNN Inference Optimization, and I/O Optimization. 
Figure~\ref{fig:optimizations} illustrates how our various optimization modules impact DeepFlame.

\subsection{Two-level Parallelization Scheme}\label{sec:parallelization_scheme}



\begin{figure}[ht]
    \centering
    \includegraphics[width=\columnwidth]{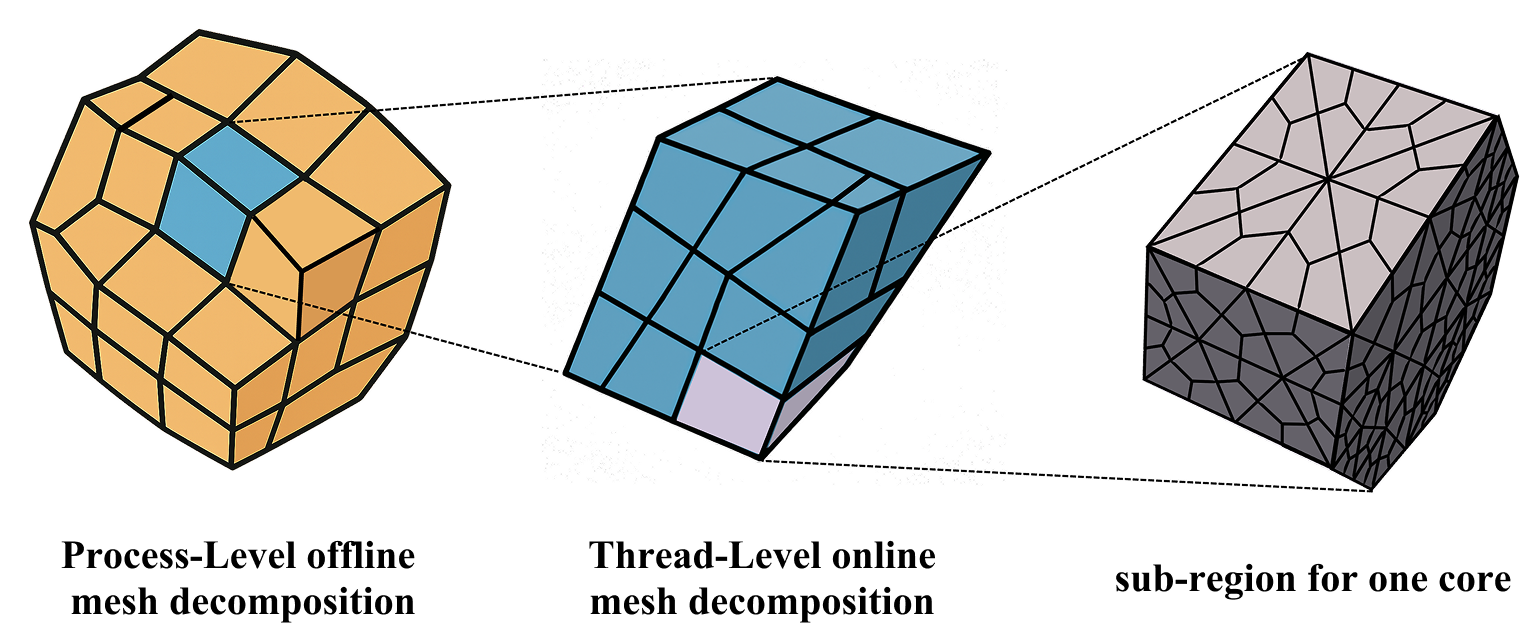}
    \caption{Two-level Parallelization Scheme.}
    \label{fig:two_level}
\end{figure}

Fig.\ref{fig:two_level} illustrates our two-level parallelization scheme: partitioning meshes into MPI processes and subsequently into thread regions, as will be detailed below.
In the optimized DeepFlame, we adopt SCOTCH as our primary mesh decomposition method due to its broad applicability, algorithmic efficiency, and ability to minimize communication overhead while maintaining load balance~\cite{pellegrini1996scotch}. 

(1) Process-level offline mesh decomposition: Previously, the mesh decomposition of Deepflame is performed offline during OpenFOAM's pre-execution phase. While the standard \textit{decomposePar} utility writes SCOTCH-generated partitions to disk (causing significant I/O bottlenecks in large-scale cases), our key contribution lies in runtime mesh refinement and a grouped parallel I/O strategy detailed in Sec.~\ref{sec:IO_optimization}.


(2) Thread-level online mesh decomposition: To address the lack of multi-threading support in OpenFOAM, we further partition the mesh within each MPI task to implement thread-level parallelization. The mesh held by individual MPI process is dynamically partitioned using SCOTCH into thread-specific sub-meshes during runtime, enabling thread-level computation. This hierarchical decomposition framework ensures load balancing for unstructured meshes, and can be further optimized together with the PDE solver as described in Sec.~\ref{sec:solver} to exploit the computational power of many-core architecture.

In the 16.2 billion-cell real-rocket-system unstructured test case, our hierarchical parallelization significantly enhances load balancing (440k mean, 459k max cell counts per process, standard deviation $\sigma$=3,222.8) , while maintaining efficient communication topology with 15 average neighbor processes and 2,855 shared faces per pair in the optimized Deepflame.

\subsection{Many-core PDE solver}\label{sec:solver} 



The DeepFlame framework solves Navier-Stokes equations through OpenFOAM using finite volume discretization with the geometric algebraic multigrid (GAMG) or preconditioned conjugate gradient (PCG, PBiStabCG) methods. A key issue is that OpenFOAM lacks multi-threading support, making it difficult to efficiently utilize the computational power of many-core architectures. While Flat-MPI parallelism incurs prohibitive memory and communication costs, hybrid process-thread parallelism requires fundamental algorithmic redesign. To address this, we developed a many-core PDE solver leveraging hierarchical mesh decomposition, achieving enhanced data locality and parallel efficiency through the following five steps.
(1) Mesh decomposition and renumbering: First we enhance the data locality of the sparse matrix by decomposing and renumbering the unstructured mesh.
(2) Customized block sparse format: Then we designed a blocked sparse format in the optimized Deepflame to further enhance data locality and data parallelism.
(3) Implementation: several key sparse operators such as SpMV, Gauss-Seidel are implemented based on the block sparse format. 
(4) Avoid write conflict. Avoid write conflicts during the parallel discretization of the PDEs. 
(5) Architecture-related optimization: Customized optimizations for Fugaku, Sunway and LS system.

Unlike library-based approaches (e.g., PETSc\cite{petsc-web-page}) that focus solely on linear algebra, our method starts from the upstream mesh structure to maximize solver performance. Note that our PDE solver can support arbitrary unstructured meshes while maintaining portability across diverse many-core architectures.

\subsubsection{Mesh decomposition and renumbering}\label{sec:mesh_decompositiona_and_renumbering}

\begin{figure}[ht]
    \centering
    \includegraphics[width=0.8\columnwidth]{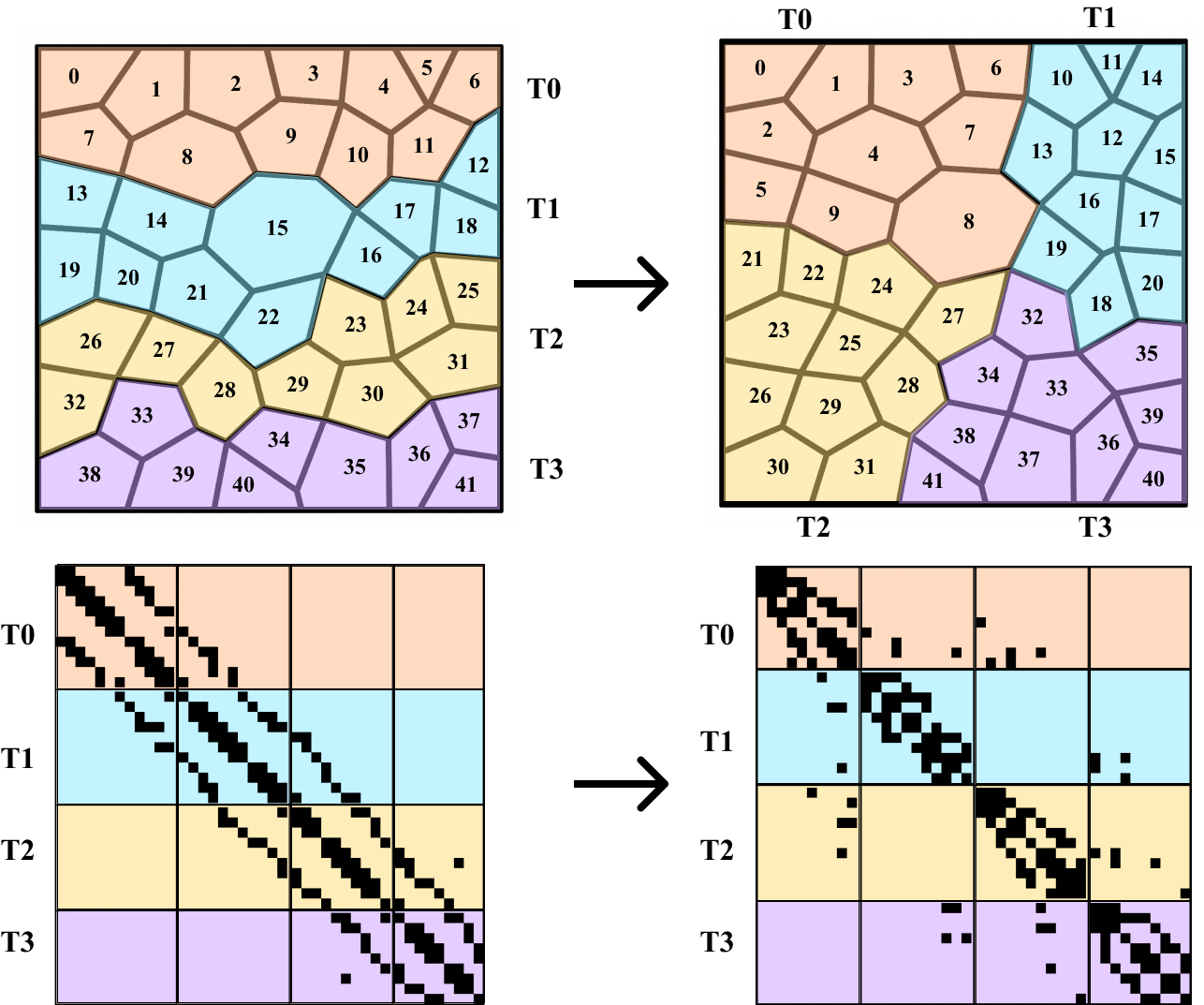}
    \caption{Thread-level mesh decomposition and renumbering.}
    \label{fig:MDAR}
\end{figure}




In this step, our optimization targets sparsity pattern restructuring through thread-level mesh decomposition and renumbering. Unstructured grids in OpenFOAM can be represented as graphs (cells→nodes, faces→edges), and the cells are then numbered to generate sparse matrix. As shown in Fig.~\ref{fig:MDAR}, cell numbering directly governs the non-zero distribution of the sparse matrix.  In our optimization, we leverage the multilevel recursive bisection algorithm in SCOTCH to minimize inter-partition edges, which directly corresponding to off-diagonal nonzeros in the sparse matrix. 
This approach aligns graph partitioning objectives with sparse matrix optimization by both concentrating nonzeros within diagonal blocks (improving data locality) and minimizing off-diagonal elements (reducing write conflicts and data dependencies among threads).

\begin{figure}[ht]
    \centering
    \includegraphics[width=0.8\columnwidth]{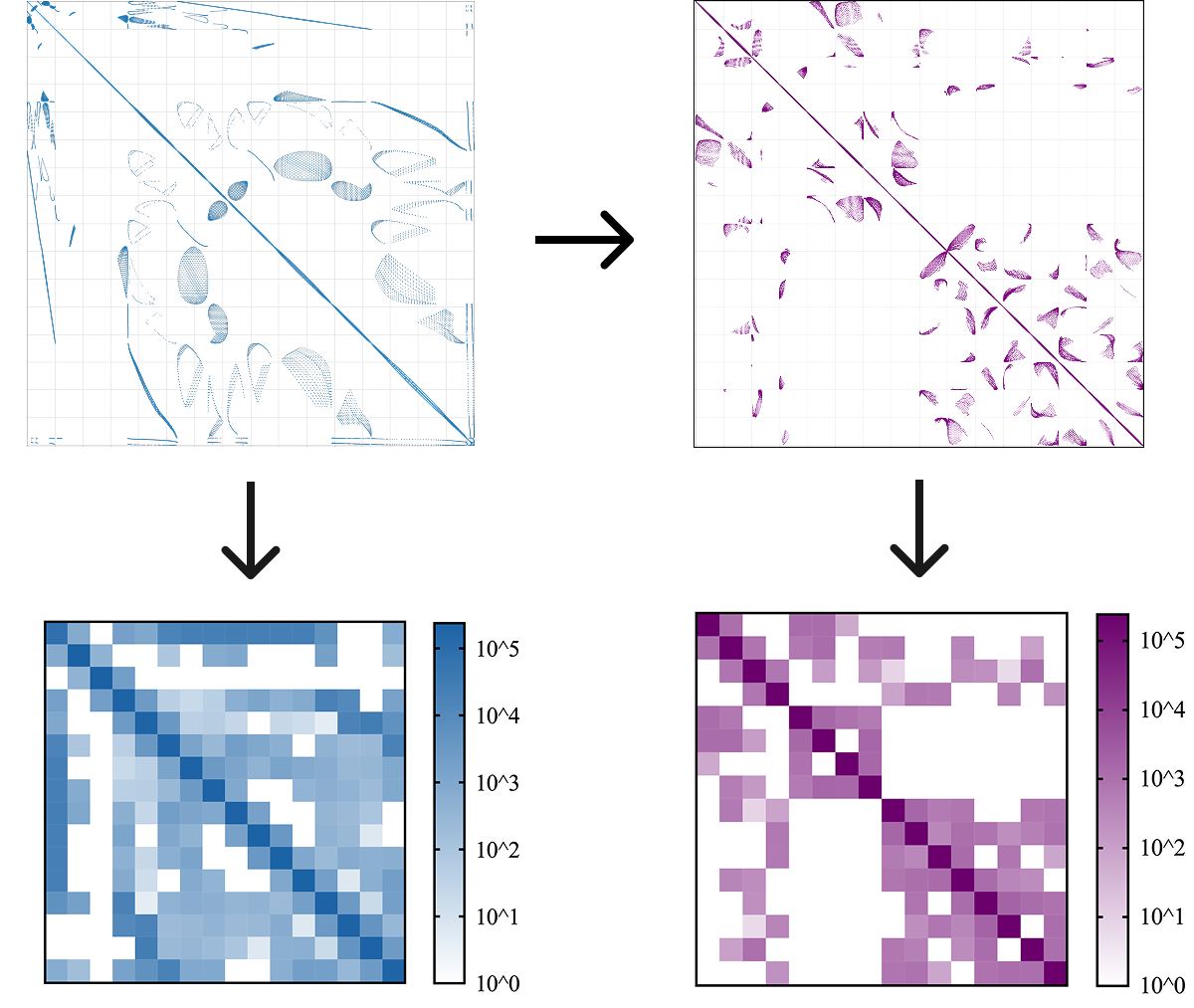}
    \caption{real rocket system decomposition and renumbering.}
    \label{fig:real_system_docomposition}
\end{figure}

Fig.~\ref{fig:MDAR} shows our thread-level decomposition and renumbering method, combining SCOTCH graph partitioning with Cuthill-McKee renumbering to structure sparse matrices into cache-efficient blocks. Each process divides its domain into $t$ subdomains, where $t$ is the thread count. Consecutive renumbering within subdomains localizes non-zero elements into diagonal blocks (thread-private computation zones), while minimizing off-diagonal entries (inter-thread communication). This induces $t\times t$ block matrices, with each threads exclusively processing one row of blocks. Fig.~\ref{fig:real_system_docomposition} quantifies the optimization impact on a real-world unstructured system:36\% fewer nonzero blocks (106→68) and 90\% reduction in off-diagonal nonzeros (16.24\%→1.63\%).

\subsubsection{Customized block sparse format}




To align with the mesh decomposition, the sparse matrix is stored in a block-wise manner. For example, when using $t$ threads, the sparse matrix is divided into $t \times t$ sub-matrices, each stored separately. While each sub-matrix could theoretically adopt a specialized sparse format based on its sparsity pattern~\cite{niu2021tilespmv}, we employ the versatile CSR format for consistency.


The new storage format introduces format conversion overhead. To address this, we implement a fast parallel algorithm converting LDU to block-sparse format.
A key observation is that during each time step, only the non-zero values change, while the sparsity pattern (matrix structure) remains static. Therefore, by precomputing positional mappings between formats, non-zero values are updated in parallel with minimal overhead. 
Tests show that the time required for our format conversion is comparable to that of a single SpMV operation.

\subsubsection{Implementation of parallel solver}\label{sec:parallel_solver}




Our hierarchical decomposition and block-structured storage enable efficient implementation of parallel PDE solvers using GAMG and PCG methods. The core computational kernels—SpMV and Gauss-Seidel smoothing—require addressing two fundamental challenges: thread-level task partitioning and resolving the Gauss-Seidel data dependencies.


Our mesh decomposition inherently resolves task partitioning through block-aligned thread assignments—each thread processes a dedicated row of matrix blocks, as shown in Fig.~\ref{fig:MDAR}. For a $t\times t$ blocked sparse matrix (where $t$ is thread count per process), each thread handles $t$ blocks, which are mostly empty due to SCOTCH-optimized sparsity patterns. Empirical validation on the real rocket system shows that our implementation can achieve effective load balancing, as the average number of non-zeros per thread is $241,634$ (max: 246198, SD: 3303.58).



Computational dependencies in Gauss-Seidel smoothing are mitigated by our decomposition strategy, which confines $98.37\%$ of nonzeros to diagonal blocks (Fig.~\ref{fig:real_system_docomposition}). The residual $1.63\%$ off-diagonal nonzeros—reduced from $16.24\%$ via SCOTCH reordering—introduce negligible cross-thread dependencies. Convergence analysis confirms these residual dependencies can be safely neglected without impacting solver stability (<$0.1\%$ residual increase per iteration).

\subsubsection{Conflict-avoid parallel matrix construction}\label{sec:parallel_construction}

Many face-to-cell operations exist in sparse system construction functions such as divergence ($\nabla\cdot \mathbf{\Psi}$), gradient ($\nabla \mathbf{\Psi}$), and Laplacian ($\Delta \mathbf{\Psi}$), among others when constructing the sparse matrix in finite volume method. These operations require updating the same cell from two faces, resulting in writing conflicts during computation. We implement a write conflict avoidance scheme based on thread-level mesh decomposition. 
As shown in Fig. \ref{fig:MDAR}, the unstructured mesh is divided into four sub-regions, with each of the four threads computing one sub-region. The edges of the grid are referred to as "faces", which are categorized into two types: intra-region faces and inter-region faces. Write conflicts can only occur when inter-region faces are computed simultaneously. Intra-region faces can be processed in parallel directly, while for inter-region faces, it is only necessary to determine the update order; write conflicts can be avoided by utilizing a multi-thread synchronization mechanism.

\subsubsection{Architecture-specific optimization}\label{sec:arch_for_pde}
\KP{
While numerous archi-tecture-specific optimizations are performed in the optimized code, we briefly introduce three key merits due to the page limitations. 
The platforms used in our work all support SIMD instructions, and important operators in the solver have been implemented with vectorization. For the programmable LDM of Sunway and the hybrid memory architecture of the LS pilot system, a double-buffering strategy is employed on both Sunway and the LS pilot systems to achieve efficient overlap of computation and data prefetching. Our solver optimization based on mesh decomposition can also naturally leverage the remote memory access mechanism among CPEs of Sunway. For instance, when computing non-diagonal blocks, data can be directly read from other CPEs to reduce memory access, or the RMA mechanism can be utilized to update edge cells of each sub-region to avoid write conflicts.
}

\subsection{DNN Inference Optimization}\label{sec:dnn}

In the baseline DeepFlame, DNN evaluation (ODENet and PRNet) consumes 61\% and 84\% of the total time on Sunway and Fugaku, respectively. We find that float-precision DNN evaluation can only reach 22.8\% and 23.4\% of peak performance on Sunway and Fugaku due to limited matrix size and sub-optimal activation function. We will focus on the DNN optimization in the following subsection.

\subsubsection{Mixed-precision computation} \label{sec:mix-dnn}

The DNN brings opportunities to use mixed precision. 
{The input of the neural network undergoes Z-score normalization to ensure the data is constrained to have a mean value of $0$ and a standard deviation of $1$.}
In our optimized DeepFlame, we replace both the weight and activation function of the DNN with FP16 after careful numerical tests. 
Compared to FP32, FP16 can reduce both memory footprint and data movement by $50\%$ and speed up the performance of the linear layer by a factor of 4.24 and 2.13 on Sunway and Fugaku, respectively. 
However, The total time for DNN inference was only reduced by 29\% due to the inefficient GeLU operation consuming most of the time. Note that except for the DNN, the other calculations such as the sparse solver and time integration are all in double precision.

\subsubsection{Tabulation for GeLU activation function} \label{sec:gelu}

The nonlinear Gaussian Error Linear Units (GeLU), which is widely adopted in GPT-3~\cite{brown2020language}, BERT~\cite{devlin2018bert}, and most other Transformers, serves as the activation function in DeepFlame. GeLU takes the form of $x\Phi(x)$, where $\Phi(x)$ is the standard Gaussian cumulative distribution function, implemented as $0.5x(1$+Tanh$(\sqrt{2/\pi}(x+0.044715x^3)))$.
However, GeLUs are inefficient on our many-core system due to mathematical transcendental functions such as Tanh. In the baseline, it accounts for 48 and 57 percent on Sunway and Fugaku, respectively. We optimize the GeLU function by exploiting the fact that $GeLU(x)=0$ when $x$ is very small, i.e., $x<-3$, and $GeLU(x)=x$ when $x$ is relatively large, i.e., $x>3$. Therefore, we approximate the GeLU function via careful 2nd-order tabulation in the range of [-3,3] with an interval of $0.01$. We remark that we have implemented both FP32- and FP16-precision tabulation in our optimized DeepFlame code.  

\subsubsection{Architecture-specific optimization}\label{sec:arch_for_dnn}


\KP{
Modern hybrid memory architectures combining small high-bandwidth memory and large low-bandwidth memory have emerged as cost-performance solutions. The core methodology employs double-buffering to parallelize computation and data transfers. By partitioning neural network inference into batched operations, systems concurrently execute current-batch computations and next-batch data prefetching. Furthermore, intermediate packed matrices from multiplication operations can be directly cached in on-package memory, eliminating redundant data movement. 
}

\subsection{I/O Optimization}\label{sec:IO_optimization}

In the CFD field, the runtime domain decomposition for unstructured mesh is particularly challenging. OpenFOAM utilizes pre-decompose mesh methods for parallel computation, leading to substantial IO challenges in large-scale simulation tasks. The \texttt{collated} data storage format is used to address the limitations of \texttt{uncollated} and \texttt{masterUncollated} formats, which generate massive files and exceed system inode limits.

When we scale our simulation to 589,824 processes, we face several challenges. First, it is nearly impossible to directly generate mesh and field data for 589,824 processes. Even if successful, it would result in terabyte-scale data files, and reading files of this size incurs significant overhead. Second, the \texttt{collated} storage format does not support parallel IO, leading to linearly increasing IO time as the simulation scale grows. Third, all processes accessing the same file simultaneously have a high overhead. Therefore, we propose multi-procedure fusion, Foam file indexing, and two-level parallel-IO to address the above issues, respectively.

\subsubsection{Runtime mesh refinement}

\begin{figure}[ht]
    \centering
    \includegraphics[width=0.9\columnwidth]{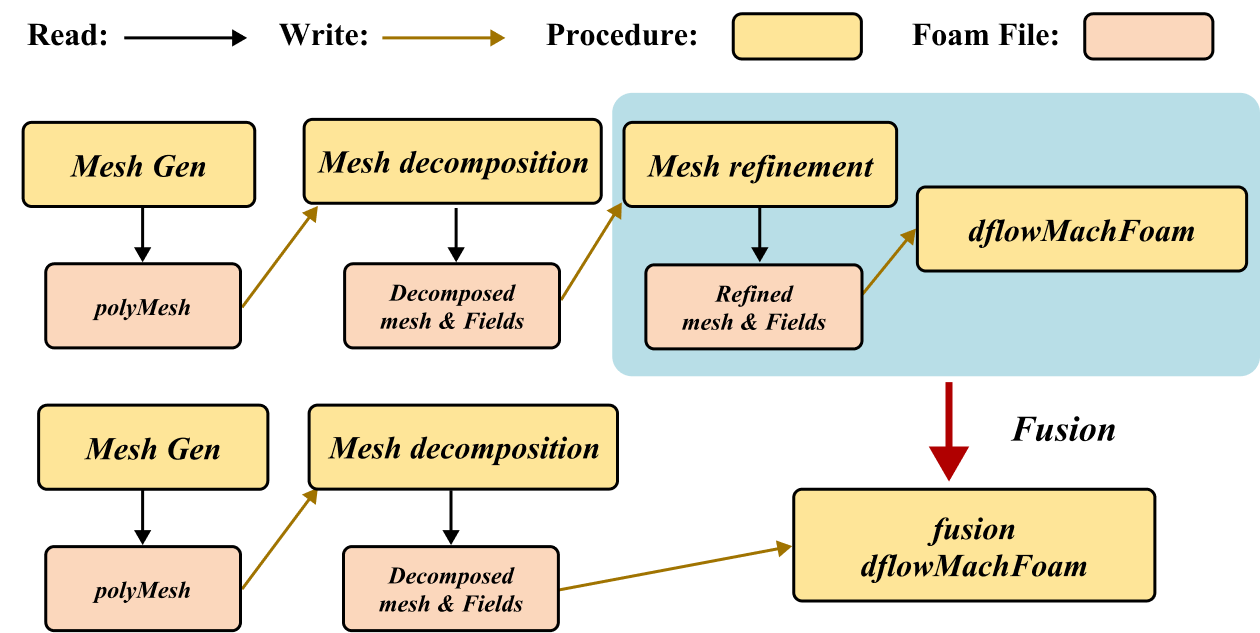}
    \caption{Runtime mesh refinement.}
    \label{fig:runtime_refinement}
\end{figure}

Our extreme-scale simulation encompasses 618 billion cells, generated through fivefold parallel refinement from an initial 19 million-cell mesh. 
However, the total size of mesh and field files for 618 billion cells can reach 121 TB (estimated), which results in a significant reading and writing overhead. To tackle this problem, we propose a \textit{runtime mesh refinement} approach, as shown in \ref{fig:runtime_refinement}. The key idea arises from the fact that parallel refinement is much faster than reading/writing TB-level files. We integrate the mesh refinement with the computation, eliminating the TB-level file read/write operation. Moreover, we only need to read the coarse mesh, reducing the input file size from 121 TB to 16~GB.

\subsubsection{Foam File Indexing}

However, even though the input file sizes are now only in GBs, 
OpenFOAM adopts the naive \textit{master read and scatter} approach, \textit{process 0} must read all the data and then distribute it to corresponding processes using \texttt{scatterv}. This is because the \texttt{collated} format in OpenFOAM does not support parallel I/O. 
To address this limitation, we develop the Foam File Indexing method to pre-generate an index file for \texttt{collated} files. This file records the start and end positions of the data needed by each process, enabling the implementation of parallel IO.
This method can be easily applied to other file formats that do not support parallel IO.

\subsubsection{Grouped Parallel I/O}


\begin{figure}[ht]
    \centering
    \includegraphics[width=0.7\columnwidth]{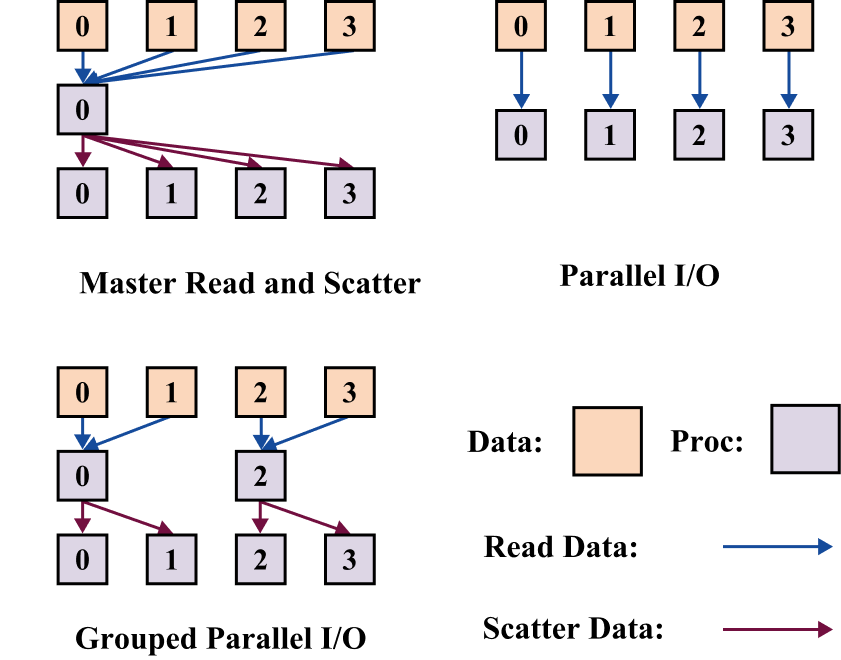}
    \caption{Grouped Parallel I/O.}
    \label{fig:grouped_parallel_IO}
\end{figure}

By now, the IO part (\textit{parallel IO}) has been optimized to enable 589,824 processes to read GB-level data in parallel. 
However, we found that when all processes read from the same file simultaneously, both the file opening time and the seek time increase linearly with the number of processes.
To address this issue, we propose \textit{Grouped parallel IO} to trade off between the process number of concurrent reading and the volume of \textit{scatter} communication, as shown in \ref{fig:grouped_parallel_IO}(e). For example, if there are \(P\) processes, we can partition these processes into \(\sqrt{P}\) groups, where each group contains \(\sqrt{P}\) processes. The first process in each group reads all the data for its group and then \textit{scatter} the data to other processes in the group. It reduces the process number of concurrent reading from \(P\) to \(\sqrt{P}\) (compared to \textit{parallel IO}), and the volume of \textit{scatter} communication from \(P\) to \(\sqrt{P}\) (compared to \textit{master read and scatter}). 

The three optimizations described in this section have resolved the long-standing IO issues that limited large-scale combustion simulations, making it possible to conduct a simulation with 618 billion cells.

\section{Physical system and HPC platform}




\subsection{Physical system used to measure performance}
The established benchmark, the 3D Taylor-Green Vortex (TGV) interacting with a diffusion flame (referred to as TGV hereafter), which has become a community standard for code validation and profiling~\cite{ABDELSAMIE2021104935,TGV,boivin2021benchmarking}, is chosen to measure the performance of the optimized DeepFlame code. We have shown in Refs.~\cite{MAO2023108842,xu2023detailed,mao2024integrated,cai2025efficient} that DeepFlame with the ODENet and PRNet models can accurately capture such multi-physical phenomena. 

The TGV system consists of a cubic computational domain, with a uniformly discretized edge length of $2\pi L$. Supercritical conditions are imposed via an initial pressure of 10~MPa, and temperature is 150~K for \ce{O2} and 300~K for \ce{CH4}. A chemical mechanism with 17~species/44~reactions~\cite{MONNIER2022111735,MONNIER20232747} is used for the LOX/\ce{CH4} combustion. 
The initial maximum velocity $u_0=$~4~m/s, giving a Reynolds number of about $Re=$~96,000 for the strong scaling case (starting case for weak scaling) with $L=$~0.48~mm. 
For the weak scalability tests, the length of $L$ in each physical direction is doubled in turn every scale up, while the mesh resolution is kept unchanged for a doubled DoF. 

\begin{figure}[htbp]
\centerline{\includegraphics[width=0.9\columnwidth]{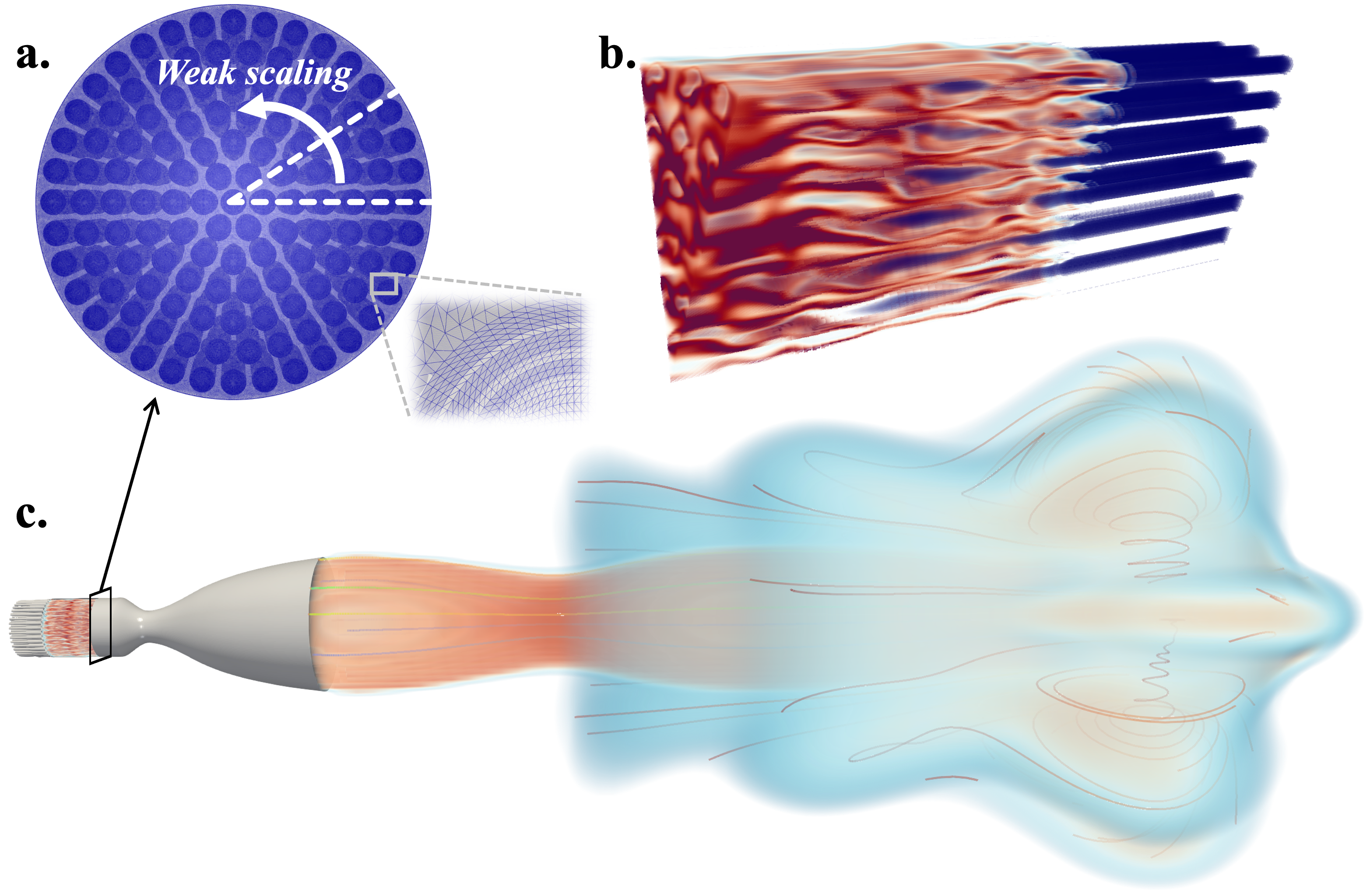}}
\caption{Liquid rocket engine simulation for weak scaling: (a) Unstructured mesh detail at the nozzle-chamber interface; (b) Flow field illustration in the single-sector baseline (smallest) test case; (c) Full-domain simulation volume rendering for temperature field.}
\label{fig:rocket}
\end{figure}

To assess DeepFlame's performance in real-world applications, we performed a full-scale rocket engine simulation. As shown in Fig.~\ref{fig:rocket}(c), this complex configuration includes 127 upstream injectors, a combustion chamber, and an exhaust nozzle, replicating actual engine operating conditions with temperatures exceeding 3000~K and pressures up to 20~MPa. The system employs the same supercritical \ce{O2}/\ce{CH4} propellants and chemical mechanism as the TGV case. The computational domain is discretized using hybrid unstructured grids totaling about 21 billion elements, with Fig.~\ref{fig:rocket}(a) illustrating mesh details at the injector-chamber interface. Notably, to maintain physical consistency across weak scaling test points, we implemented a sector-based domain decomposition strategy where computational size increases through angular sector sweeping. Fig.~\ref{fig:rocket}(b) displays results from the single-sector ($\angle 22.5^\circ$) configuration, while the full-size domain represents the largest weak scaling test case.
The ODENet model is of the size (20, 2048, 4096, 2048, 1024, 512, 17), and the PRNet consists of a model of the size (3, 1024, 512, 256, 1) for density and a model of the size (3, 2048, 1024, 512, 4) for the temperature and other transport properties. To test the performance, the DeepFlame equations with the above NN models were numerically advanced for 100 time-steps. 

\subsection{HPC systems and software environment}

All our tests were carried out on three many-core platform : Sunway, Fugaku and the LS pilot system.

The new Sunway is a many-core heterogeneous supercomputer equipped with $102,400$ computing nodes. Each node is powered by a sw26010-pro CPU and interconnected through a $16$:$3$ ($256$:$48$) oversubscribed multi-layer fat-tree network. Each sw26010-pro CPU achieves theoretical peak performances of 13.824 TFlop/s in double precision and 55.296 TFlop/s in FP16 precision. 

Fugaku supercomputer, housed at the RIKEN Center for Computational Science, is currently ranked No. 6 on the Top500 list~\cite{top500}. It comprises 158,976 computing nodes, each equipped with an A64FX CPU \cite{sato2020co}, and boasts a peak performance of 537 PFlop/s in double precision and 2.1 EFlop/s in FP16 precision. 

\KP{
The LS pilot system comprises a 256-node cluster architecture, with each node containing two LX2 high-performance CPUs that collectively provide over 256 processing cores. 
The LX2 CPU employs a system-on-chip design with dual computing dies integrated in one package. Each die features 128GB off-die DDR memory divided into 4 NUMA domains, enhanced by a System DMA interface for optimized data transfers between DDR and on-package memory. Its core architecture supports vector and matrix engine, including double-precision FP SIMD instructions and native 8x8 matrix operations in the execution pipeline.
}

A two-level parallelism scheme of processes and threads is employed on all three machines. On Sunway, the MPI+athread programming model is utilized, with each process controlling one core group. On the Fugaku and the LS pilot system, the MPI+OpenMP programming model is adopted, with each process managing one NUMA domain.

\subsection{Measurement Methodology}

The total ﬂoating point operations (FLOPs) of the performed calculations are collected via counting the effective FLOPs during neural network inference and sparse matrix linear equations solving, which is less than the actual FLOPs executed during the entire DeepFlame execution. The following criteria are used to measure the performance of our program.

\begin{itemize}

\item \textbf{Time-to-solution}, defined as $\frac{\textsf{DeepFlame loop time}}{\textsf{DoF} \times \textsf{flow-cycle per loop}} $. The ``DeepFlame loop time'' is the wall-clock time elapsed for a single time step, and the ``flow-cycle'' refers to a characteristic physical flow time, giving an overall unit of [s/DoF/cycle].

\item \textbf{Peak performance}, defined as $\frac{ \textsf{total FLOPs}} {\textsf{DeepFlame loop time}}$.


\end{itemize}

\section{Performance Results} \label{sec:perfRes}



\subsection{Accuracy Test}


\begin{figure}[htbp]
\centerline{\includegraphics[width=\columnwidth]{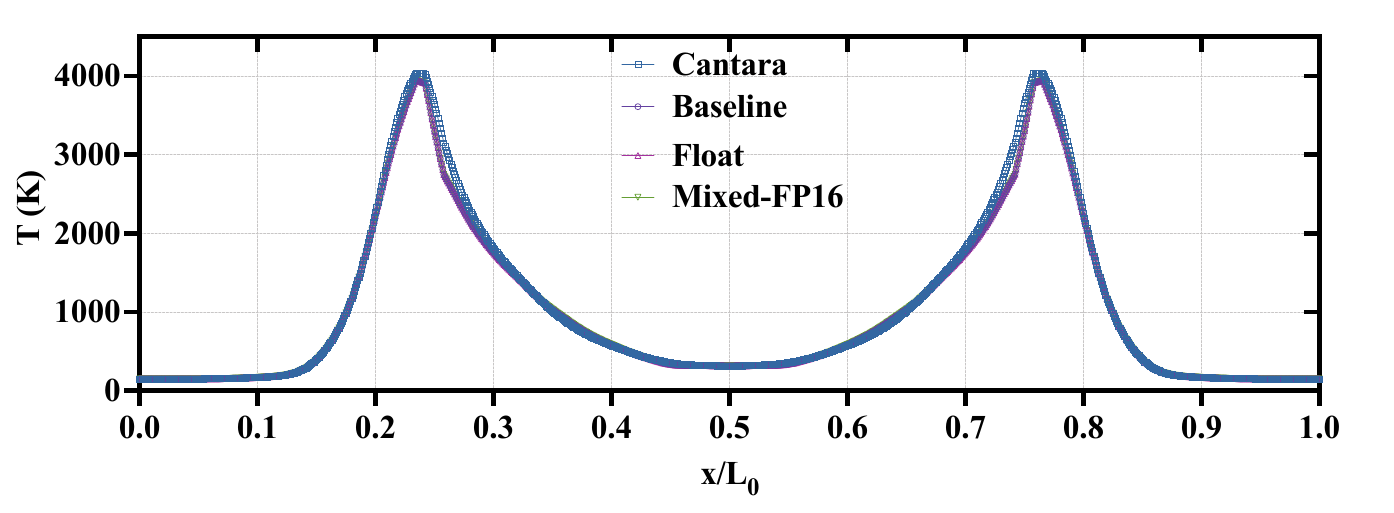}}
\caption{Comparison of temperatures gained from ordinary method and inference with different precision. "Cantara" denotes the traditional ODE solver method, validated for computational accuracy. "Baseline" represents results from the unoptimized DeepFrame software. Both "Float" and "Mixed-FP16" employ manually implemented DNN inference: "Float" uses float precision with a float-precision lookup table fitting approach for GeLU, whereas "Mixed-FP16" applies half-precision.}
\label{fig:accuracy}
\end{figure}

The numerical optimizations for DNN architecture in Sections~\ref{sec:mix-dnn} and \ref{sec:gelu}, while computationally efficient, could potentially affect numerical precision. We validated accuracy against the original DeepFlame implementation through rigorous testing. As shown in Fig.\ref{fig:accuracy} and Table.\ref{tab:precisions}, our optimized float-precision and mixed-fp16 variants demonstrate maximum relative errors of 1.49\% and 1.51\% compared to the reference, with absolute errors constrained within 62.2 and 64.2 units respectively across all test cases. These results verify the numerical stability of our approaches.

\begin{table}[]
\centering
\caption{Simulation errors with different precisions.}\label{tab:precisions}
\begin{tabular}{c|cc|cc}
\toprule[1pt]
       & \multicolumn{2}{c|}{relative error [\%]}            & \multicolumn{2}{c}{absolute error [K]}            \\ \cline{2-5}
       & avg.                 & max.                  & avg.                 & max.                 \\ \hline
Float &   0.28\%            &   1.49\%             &    1.91                  &     62.2                 \\
Mixed-FP16 & 0.29\% & 1.51\% & 1.96 &64.2 \\ 
\bottomrule[1pt]
\end{tabular}
\end{table}

\subsection{Step-by-step performance improvement} \label{sec:step-by-step-test}

\begin{figure}
    \centering
    \includegraphics[width=1\columnwidth]{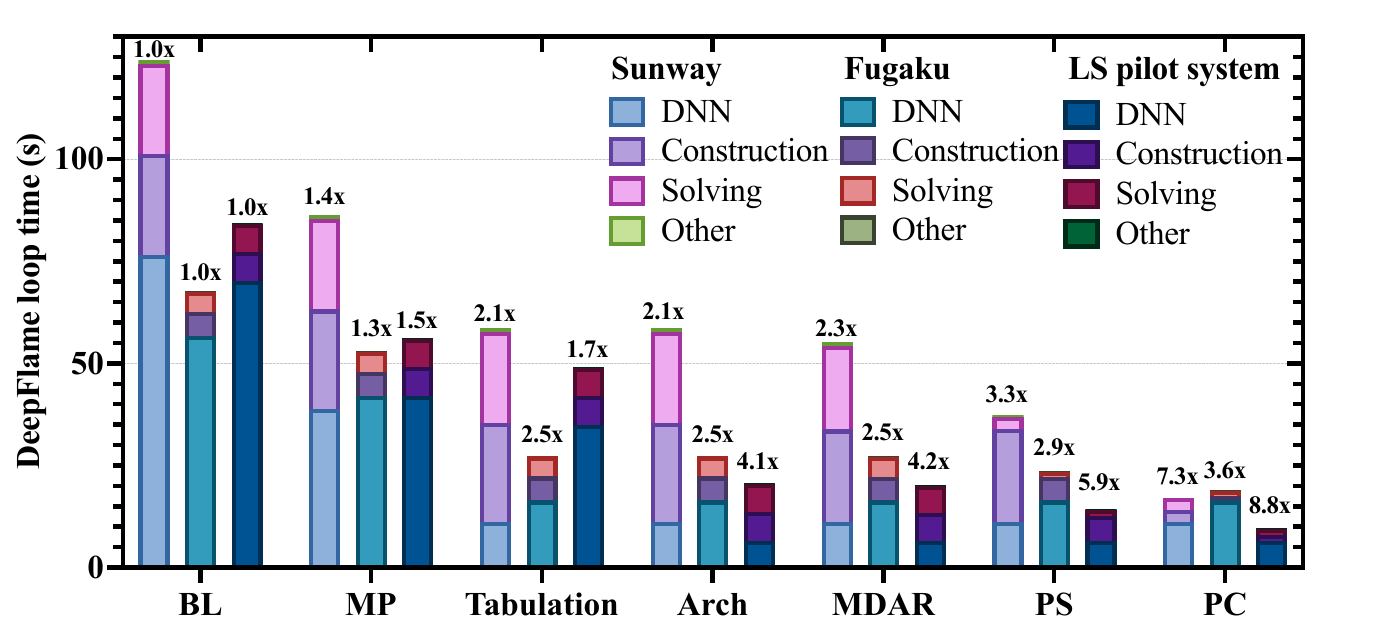}
    \caption{\KP{Step-by-step performance improvement of DeepFlame with $25,165,824$ cells on Sunway, Fugaku and the LS pilot system, respectively.}}
    \label{fig:step-by-step}
\end{figure}


Fig.~\ref{fig:step-by-step} illustrates performance optimizations of our approach across three many-core architectures using a 25,165,824-cell TGV system. 
The DeepFlame loop time comprises four components: DNN inference, Construction, Solving, and Others. The "DNN" time encompasses both ODENet and PRNet execution, while "Construction" and "Solving" encompass the construction and solution of all sparse systems, respectively.
Key optimizations are denoted as: BL (Baseline), MP (Mixed-precision, Sec.\ref{sec:mix-dnn}), Tabulation (GeLU approximation, Sec.\ref{sec:gelu}), Arch (Architecture-specific tuning, Sec.\ref{sec:arch_for_dnn}), MDAR (Mesh Decomposition and Renumbering, Sec.\ref{sec:mesh_decompositiona_and_renumbering}), PS (Parallel Solver, Sec.\ref{sec:parallel_solver}), and PC (Parallel Construction, Sec.\ref{sec:parallel_construction}).
Our custom DNN implementation (without third-party frameworks) initially employs float precision in the baseline, utilizing optimized BLAS libraries for fully-connected layers and tanh-based GeLU activation. The baseline already employs optimized BLAS routines with full multi-threaded core utilization through OpenMP parallelization.

\subsubsection{DNN Inference module}

\KP{
For the DNN inference module, our optimizations achieve 6.9x, 3.4x, and 10.6x speedups on Sunway, Fugaku, and the LS pilot system respectively, while achieving simulation speedups of 2.1x, 2.4x, and 4.6x. The higher acceleration on the LS pilot system stems from our architecture-specific optimizations (e.g., hybrid memory architecture and matrix computation units) harnessing its powerful AI capabilities. The module reaches peak computational efficiencies of 40.0\%, 40.2\%, and 42.6\% across these systems respectively.
}

\subsubsection{PDE solving module}
\KP{
For the PDE solving module, our three optimizations provide improvements of 7.8x, 4.6x, and 4.7x for Sunway, Fugaku, and the LS pilot system, respectively, and the simulation speed are improved by 3.42x, 1.4x, and 2.1x, respectively. The optimization achieves the most significant improvement on Sunway due to its higher number of threads and additional architecture-specific optimizations, such as the use of double buffering and RMA mechanisms. In fact, the PDE solving module on Fugaku demonstrates the best performance owing to its high memory bandwidth. The hybrid memory architecture of the LS pilot system greatly enhances its memory bandwidth but also lead to higher optimization costs.
}

\subsubsection{Comparison between three many-core system}
\KP{
Our optimizations achieve speedups of 7.3x, 3.6x, and 8.8x across the three systems respectively. Post-optimization, the DNN inference module accounts for 64.9\%, 87.4\%, and 68.9\% of computational workload, while PDE solving module constitutes 35.0\%, 12.2\%, and 30.3\% respectively. The systems demonstrate computational efficiencies of 28.5\%, 35.1\%, and 29.4\%. Notably, on Fugaku, PDE solving module exhibits the smallest proportion yet highest floating-point efficiency due to its low performance-to-bandwidth ratio. Both Sunway and the LS pilot system leverage superior floating-point capabilities, with the latter's hybrid memory architecture mitigating bandwidth constraints. Their enhanced half-precision computation performance enables faster time-to-solution compared to conventional architectures.
}

\subsection{Comparison between structured and unstructured meshes}

\begin{figure}
    \centering
    \includegraphics[width=\columnwidth]{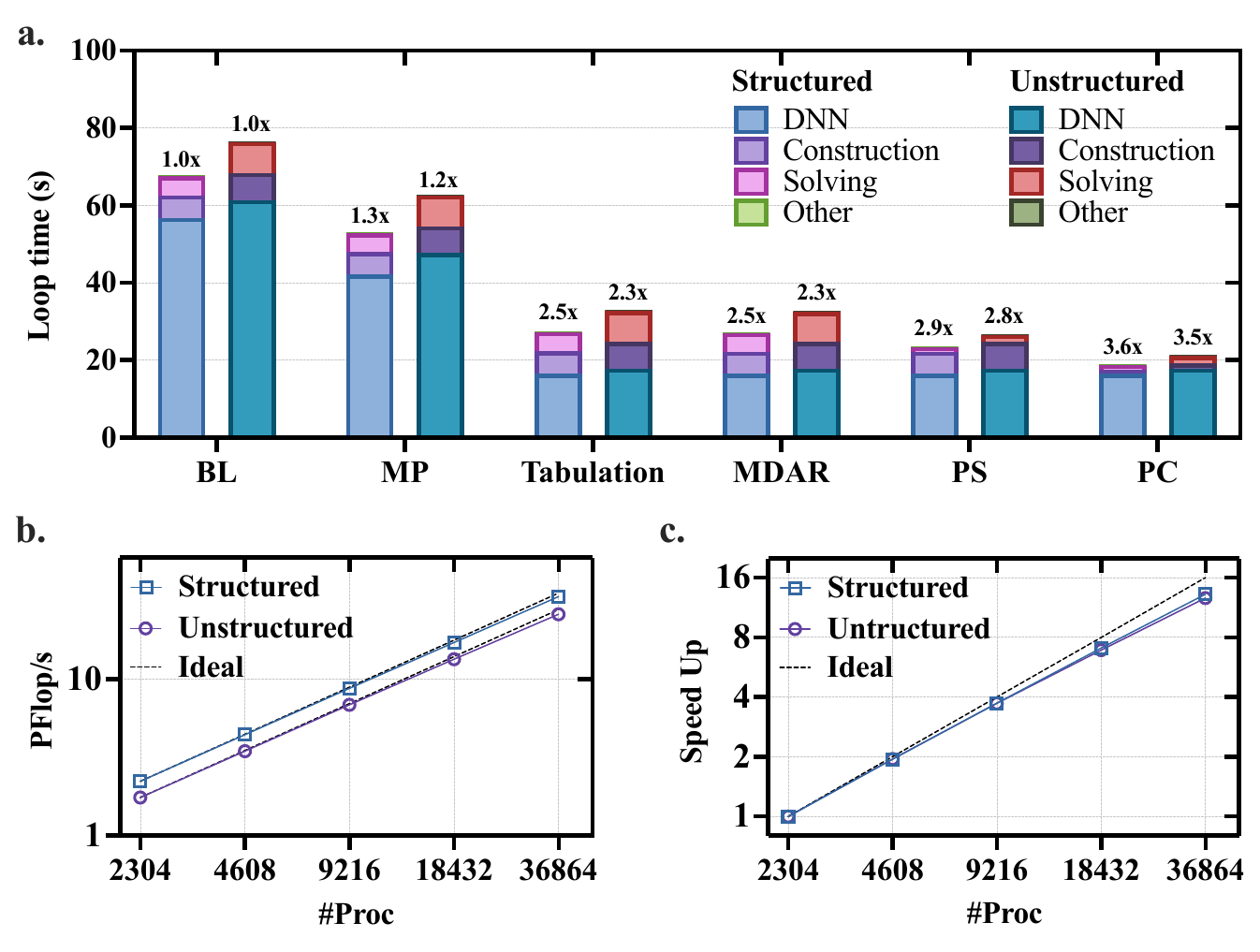}
    \caption{Performance comparison between structured and unstructured meshes. (a). Step-by-step performance improvement comparison. (b). Weak scaling comparison. (c). Strong scaling comparison. }
    \label{fig:structured_unstructured}
\end{figure}


This section compares structured vs. unstructured grid performance on Fugaku under identical hardware and optimization settings (non-grid-specific algorithms). Additionally, since the two systems have a similar number of cells, both test cases occupy approximately 70-75\% of the memory space. Structured grid cases employ the TGV benchmark with two-level \textit{simple} decomposition, while unstructured grid cases use our real rocket system with two-level \textit{scotch} decomposition.



Fig.~\ref{fig:structured_unstructured}(a) demonstrates our method's optimization performance on structured versus unstructured grids, yielding 3.58x and 3.50x speedups with 35.1\% and 32.2\% half-precision computational efficiency, respectively. This efficiency gap stems from two factors: (1) The unstructured grid experiences slight load imbalance, with average and maximum cell counts per process being 561,496 and 567,053 compared to structured grids' uniform 524,288 cells; (2) Structured grids generate sparse matrices with superior data locality.



Fig.~\ref{fig:structured_unstructured}(b) and Fig.~\ref{fig:structured_unstructured}(c) show the weak and strong scaling performance for structured and unstructured grid simulations, respectively. When scaling to 16x processes, weak scaling efficiencies reach 94.9\% (structured) and 93.1\% (unstructured), while strong scaling efficiencies attain 82.5\% and 79.0\%, respectively. DeepFlame's communication involves two primary components: global \textit{Allreduce} operations in the conjugate gradient solver and halo exchanges from domain decomposition. While no distinction exists between grid types for the former, the latter reveals notable differences - unstructured grids require communication with 15 neighboring processes on average versus 6 for structured grids. Nevertheless, these constant-factor variations exert limited influence on overall scalability.

\subsection{Strong Scaling}\label{sec:strong_scaling}

\begin{figure}[htbp]
\centerline{\includegraphics[width=\columnwidth]{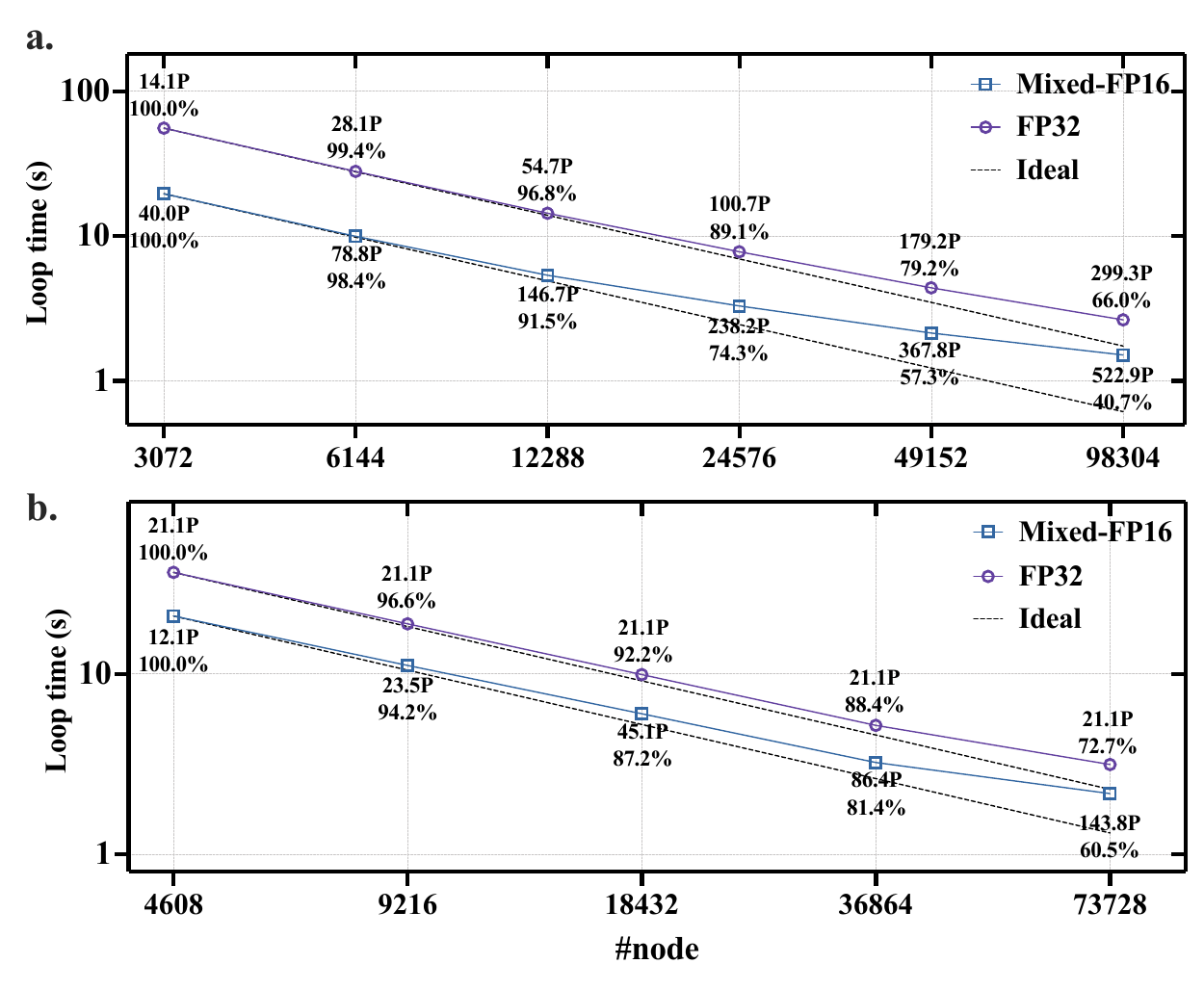}}
\caption{
Strong scaling is evaluated for two configurations: (a) the TGV benchmark with 19.3 billion cells on Sunway (3,072–98,304 nodes) and (b) the 9.7 billion-cell CH$_4$ system on Fugaku (4,608–73,728 nodes). Both experiments report peak computational performance in PFlop/s and demonstrate corresponding parallel efficiency metrics.
}
\label{fig:strong_scaling}
\end{figure}

Fig.~\ref{fig:strong_scaling} shows the strong scaling of TGV benchmark of Sunway and Fugaku. The system sizes are 19,327,352,832 cells and 9,663,676,416 cells, which are inaccessible with the original code. 
Both scaling tests on Sunway and Fugaku start from 18,432 MPI tasks.

On Sunway, the optimized DeepFlame scales well to 98,304 computing nodes (98\% of the entire machine). The parallel efficiency is 40.7\% in the mixed-FP16 and 66.0\% in FP32 precision by setting the performance with 3,072 computing nodes as a baseline. When scaling to 98,304 computing nodes, the optimized DeepFlame can reach 522.9 PFlop/s and 299.3 PFlop/s in mixed-FP16 and FP32 precision, respectively. The corresponding time-to-solution of one-time step of flame simulation with detailed transport and chemistry accuracy can reach $\mathbf{2.7 \times 10^{-9}}~{\rm \mathbf{s/DoF/cycle}}$ with time step set to 10 nanoseconds. 

The optimized DeepFlame scales up to 73,728 computing nodes on Fugaku with a parallel efficiency of 60.5\% in mixed-FP16 and 72.7\% in FP32 precision by setting the performance with 4,608 computing nodes as baseline. The peak performance reaches 208.6 PFlop/s in mixed precision and 143.8 PFlop/s in fp32 precision on 73,728 nodes. The corresponding time-to-solution reach 7.7 $\times$ 10$^{-9}$~{\rm s/DoF/cycle} with time step set to 10 nanoseconds. 

\subsection{Weak Scaling}\label{sec:weak_scaling}

\begin{figure}[htbp]
\centerline{\includegraphics[width=\columnwidth]{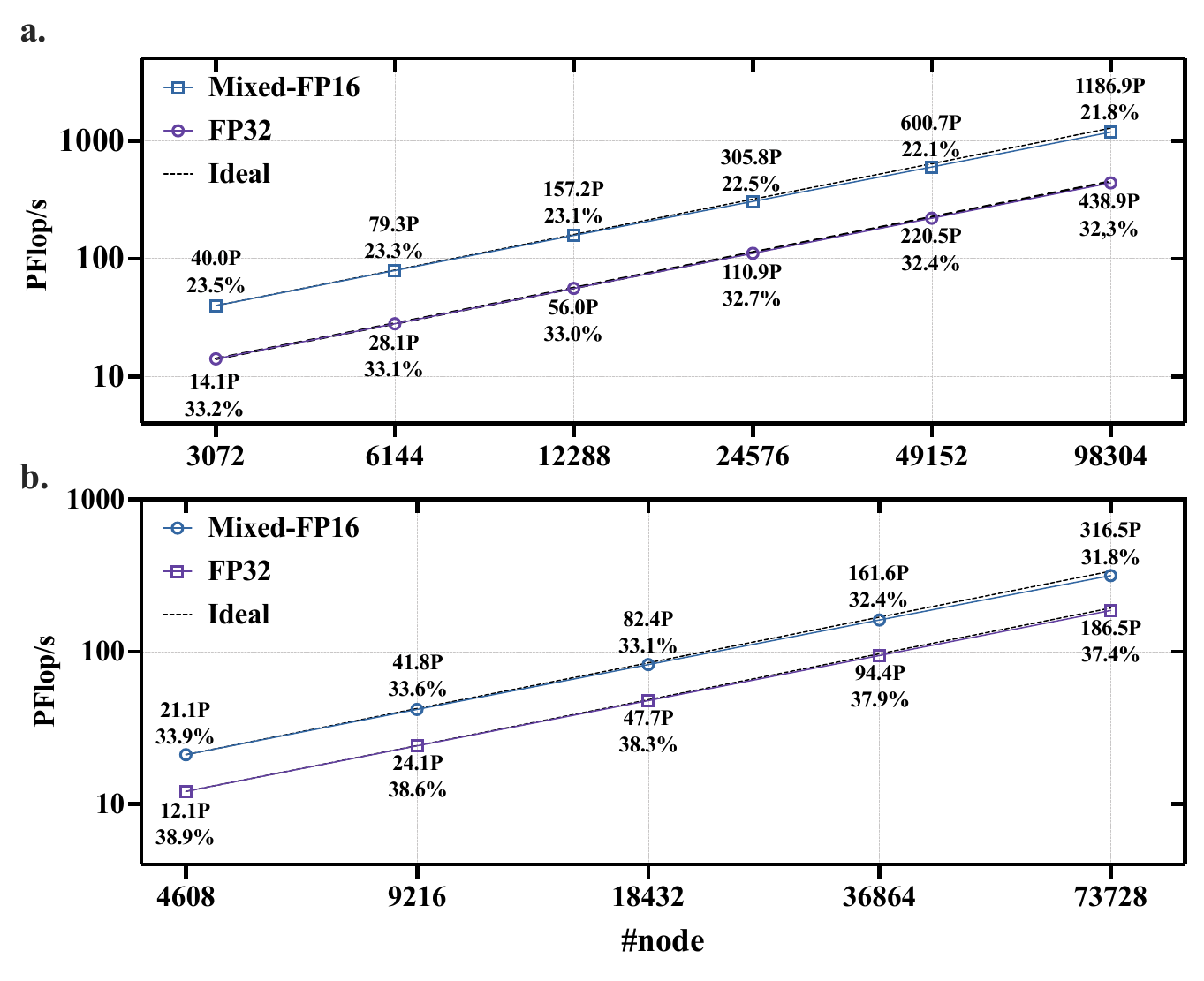}}
\caption{Weak scaling tests were conducted for the TGV benchmark: (a) scaling from 19,327,352,832 to 618,475,290,624 cells on Sunway; (b) scaling from 9,663,676,416 to 154,618,822,656 cells on Fugaku. The achieved peak Flop/s and corresponding percentage of theoretical peak performance are reported.}
\label{fig:weak_scaling}
\end{figure}

The weak scaling of the optimized DeepFlame is measured in terms of the system size and Flop/s for the TGV benchmark on Sunway and Fugaku. Fig.~\ref{fig:weak_scaling} shows near-perfect weak scaling with respect to the number of computing nodes for the FP32 and mixed-FP16 precision. The corresponding physical system reaches up to \textbf{618 billion} cells on Sunway and 154 billion on Fugaku. Note that Sunway can accommodate a bigger system due to the bigger memory capacity. 

On Sunway, the optimized DeepFlame attains a parallel efficiency of 92.74\% in mixed-FP16 precision and 97.31\% in FP32 precision when scaling from 3,072 to 98,304 computing nodes. The peak performance is \textbf{1.18 EFlop/s} (21.8\% of the theoretical peak) in mixed-FP16 and 438.9 PFlop/s (32.3\% of the theoretical peak) in FP32 precision on 98,304 nodes (98\% of the entire machine). On Fugaku, the optimized DeepFlame scales from 4,608 to 73,728 computing nodes, reaching a parallel efficiency of 93.59\% in mixed-FP16 and 96.2\% in FP32 precision. The peak performance reaches 316.5 PFlop/s (31.8\% of theoretical peak) in mixed-FP16 precision and 186.5 PFlop/s (37.4\% of theoretical peak) in FP32 precision on 73,728 nodes (half of the entire machine).

The time-to-solution reaches $1.2\times10^{-9}~{\rm s/DoF/cycle}$ on $98,304$ computing nodes of Sunway. This is, as far as we know, at least 10,000 times faster compared to the current state-of-the-art for supercritical flame simulation at detailed transport and chemistry accuracy.

\section{Conclusion}

This paper introduces optimizations for deep learning-based supercritical flame simulation software, DeepFlame, while maintaining real-fluid physics and chemical accuracy. Our analysis identifies three computational bottlenecks hindering DeepFlame's efficiency and scalability on exascale many-core systems, and proposes four contributions to resolve them.

First, a two-level parallelism scheme addresses the inability to utilize modern many-core supercomputers, enabling efficient computing on million-core architectures.
Second, computational optimizations for both DNN inference and PDE solving modules maximize floating-point performance, particularly through a mesh decomposition-based PDE solver that effectively addresses issues of poor locality, low computational density, write conflicts, and dependency constraints. Third, three I/O optimization strategies overcome bottlenecks in ultra-large-scale unstructured mesh combustion simulations.

The optimized code achieves 1.18 EFlop/s (21.8\%) mixed-FP16 precision on Sunway and 316.5 PFlop/s (31.8\%) on Fugaku. It enables combustion simulations with 618 billion cells, breaking previous spatiotemporal scale limitations while maintaining real-fluid transport and chemical accuracy. These advancements establish high-fidelity supercritical flame simulations as predictive tools for next-generation rocket engines and ultra-high energy density propulsion systems.

\begin{acks}
This work is supported by the following funding: National Science Foundation of China (92270203, 92270206, T2125013, 62372435, 62032023, 61972377, 61972380, \ T2293702, \ 523B2062, \ 52441603, \ 52276096). 
The AI-driven experiments, simulations and model training were performed on the robotic AI-Scientist platform of Chinese Academy of Sciences.
\end{acks}

\bibliographystyle{ACM-Reference-Format}
\bibliography{Reference}


\end{document}